\documentclass[aps,prb,twocolumn,showpacs]{revtex4}

\usepackage{graphicx}
\usepackage{amsmath}
\usepackage{amssymb}

\newcommand{\os}[2]{#1_{#2\,\sigma}}
\newcommand{\osd}[2]{#1_{#2\,\sigma}^\dagger}

\newcommand{\ou}[2]{#1_{#2\,\uparrow}}
\newcommand{\oud}[2]{#1_{#2\,\uparrow}^\dagger}

\newcommand{\od}[2]{#1_{#2\,\downarrow}}
\newcommand{\odd}[2]{#1_{#2\,\downarrow}^\dagger}

\newcommand{\vv}[1]{{\bf{#1}}}

\newcommand{\Ek}{E_{{\rm{h}}\vv{k}}}
\newcommand{\Ekq}{E_{{\rm{h}}\vv{k+q}}}

\newcommand{\xik}{\bar{\xi}_{\vv{k}}}
\newcommand{\xikq}{\bar{\xi}_{\vv{k+q}}}

\newcommand{\Dk}{\bar{\Delta}_{{\rm{hZ}}}(\vv{k})}
\newcommand{\Dkq}{\bar{\Delta}_{{\rm{hZ}}}(\vv{k+q})}

\newcommand{\nF}{n_{\rm{F}}}

\begin{document}

\title{Electromagnetic response in kinetic energy driven cuprate
superconductors: Linear response approach}

\author{Mateusz Krzyzosiak}
\affiliation{Department of Physics, Beijing Normal University,
Beijing 100875, China} \affiliation{Institute of Physics, Wroc{\l}aw
University of Technology, Wybrze\.{z}e Wyspia\'{n}skiego 27, 50-370
Wroc{\l}aw, Poland}

\author{Zheyu Huang and Shiping Feng$^{*}$}

\affiliation{Department of Physics, Beijing Normal University,
Beijing 100875, China}

\author{Ryszard Gonczarek}
\affiliation{Institute of Physics, Wroc{\l}aw University of
Technology, Wybrze\.{z}e Wyspia\'{n}skiego 27, 50-370 Wroc{\l}aw,
Poland}

\begin{abstract}
Within the framework of the kinetic energy driven superconductivity,
the electromagnetic response in cuprate superconductors is studied
in the linear response approach. The kernel of the response function
is evaluated and employed to calculate the local magnetic field
profile, the magnetic field penetration depth, and the superfluid
density, based on the specular reflection model for a purely
transverse vector potential. It is shown that the low temperature
magnetic field profile follows an exponential decay at the surface,
while the magnetic field penetration depth depends linearly on
temperature, except for the strong deviation from the linear
characteristics at extremely low temperatures. The superfluid
density is found to decrease linearly with decreasing doping
concentration in the underdoped regime. The problem of gauge
invariance is addressed and an approximation for the dressed current
vertex, which does not violate local charge conservation is proposed
and discussed.
\end{abstract}


\maketitle

\section{Introduction}

Observation of superconductor's response to a weak external
electromagnetic stimulus allows us to collect a number of subtle
characteristics \cite{schrieffer83}. The way the magnetic field is
expelled from a superconducting (SC) sample in the spectacular
Meissner effect can be used to infer about many fundamental features
of the system. Therefore the phenomena at the length scale of the
magnetic field penetration depth $\lambda$, i.e. in the region at
the edge of the sample where the induced supercurrents effectively
screen the external magnetic field, are subject to intensive studies
both on the theoretical and the experimental fronts of the research
in cuprate superconductors \cite{bonn96,tsuei00}. In particular, the
magnetic field penetration depth can be used as a probe of the
pairing symmetry since it can distinguish between a fully gapped and
a nodal quasiparticle excitation spectrum \cite{bonn96,tsuei00}. The
former results in the thermally activated (exponential) temperature
dependence of the magnetic field penetration depth, whereas the
latter one implies a power law behavior.

The magnetic field penetration depth is a basic parameter of
superconductors, closely related to the superfluid density
\cite{schrieffer83}. Earlier on, the linear temperature dependence
of the magnetic field penetration depth $\lambda(T)$ was observed
for the cuprate superconductor YBa$_2$Cu$_3$O$_{7-y}$ at low
temperatures ($T=4$K$\sim$20K) \cite{hardy93}, which first provided
a strong experimental support for the nodes in the d-wave SC gap
function of cuprate superconductors, then confirmed by the
angle-resolved photoemission spectroscopy (ARPES) experiments
\cite{ding9495,damascelli03}. Later, this linear temperature
dependence of the magnetic field penetration depth has been observed
in different families of cuprate superconductors
\cite{kamal98,jackson00,panagopoulos99,pereg07}. However, at
extremely low temperatures ($T<4$K), the linear temperature
dependence of the magnetic field penetration depth is modified, and
a nonlinearity emerges \cite{khasanov04,suter04,sonier99}. Moreover,
some indications of nonlocal effects giving rise to the nonlinearity
have been reported in the field dependence of the effective magnetic
field penetration depth in cuprate superconductors \cite{sonier99}.
Furthermore, the doping dependence of the electromagnetic response
in cuprate superconductors has been studied in terms of the
zero-temperature superfluid density. The superfluid density is
proportional to the squared amplitude of the coherent macroscopic
wave function describing the SC charge carriers, and therefore it is
an important physical quantity and can provide significant
information about the SC state. In particular, the superfluid
density of cuprate superconductors in the underdoped regime vanishes
more or less linearly with decreasing doping concentration
\cite{uemura8991,broun07,bernhard01}. This in turn gives rise to the
linear relation between the critical temperature $T_{\rm{c}}$ and
the superfluid density observed in the underdoped regime
\cite{uemura8991}.

Theoretically, the electromagnetic response in cuprate
superconductors has been extensively studied based on the the
phenomenological Bardeen-Cooper-Schrieffer (BCS) formalism with the
d-wave SC gap function \cite{yip92,kosztin97,franz97,li00,sheehy04}.
It has been shown \cite{tsuei00,kosztin97} that for a d-wave
superconductor in the local limit ($\zeta\ll\lambda$, where $\zeta$
is the coherence length), the simple d-wave pairing state (assuming
a tetragonal symmetry and ignoring the dispersion in the c-axis
direction) gives the magnetic field penetration depth
$\lambda(T)\propto T/\Delta_{0}$, where $\Delta_{0}$ is the
zero-temperature value of the d-wave gap amplitude. In particular,
it has been argued that this linear temperature dependence of the
magnetic field penetration depth is attributed to the excitation of
quasiparticles out of the condensate at the nodes of the SC gap
function. Furthermore, the fact
\cite{kamal98,jackson00,panagopoulos99,pereg07} that this linear
relation holds down to very low doping concentrations suggests that
near the nodes these quasiparticle excitations are well described by
a simple BCS-like formalism with the d-wave SC gap function, even
for the doping concentration $\delta\to 0$ \cite{sheehy04}. This is
also consistent with the ARPES experiments \cite{matsui}. However,
this depends sensitively on the quasiparticle scattering. In
particular, at extremely low temperatures, the coherence length may
diverge at the nodes. This may imply that the local condition no
longer holds, and the electromagnetic field varies significantly
over the size of a Cooper pair. Consequently, the nonlocal effect
emerges \cite{suter04} and then plays an important role in the
electromagnetic response of cuprate superconductors
\cite{yip92,kosztin97,franz97,li00,sheehy04}. It has been suggested
\cite{yip92,kosztin97,franz97,li00,sheehy04} that nonlocal effects
can imply a crossover from the linear temperature dependence of the
magnetic field penetration depth at low temperatures to a nonlinear
one in the extremely low temperature range. To the best of our
knowledge, the electromagnetic response in cuprate superconductors
has not been treated starting from a microscopic SC theory, and no
explicit calculations of the doping dependence of the superfluid
density in the underdoped regime have been made so far.

Recently, a kinetic energy driven SC mechanism has been developed
\cite{feng0306}, where the charge carrier-spin interaction from the
kinetic energy term induces a charge carrier pairing state with the
d-wave symmetry by exchanging spin excitations. Then the electron
Cooper pairs originating from the charge carrier pairing state are
due to charge-spin recombination, and their condensation reveals the
d-wave SC ground-state. In particular, this SC-state is controlled
by both the SC gap function and the quasiparticle coherence, then
the maximal SC transition temperature occurs around the optimal
doping, and decreases in both underdoped and overdoped regimes. The
unique feature of this kinetic energy driven SC mechanism is that
the pairing comes out from the kinetic energy by exchanging spin
excitations and is not driven by the magnetic superexchange
interaction as in the resonant valence bond type theories
\cite{anderson87}. Within the framework of the kinetic energy driven
superconductivity, we have discussed the low energy electronic
structure \cite{feng07,guo07} of cuprate superconductors and the
spin response \cite{feng0306,cheng08}, and qualitatively reproduced
some main features of ARPES experiments \cite{ding9495,damascelli03}
and inelastic neutron scattering \cite{dai01,arai99} measurements on
cuprate superconductors.

The layered crystal structure gives rise to a strong anisotropy of
cuprate superconductors, and it is possible to observe both in-plane
and inter-plane electromagnetic responses. The former one is
characterized by the ab-plane magnetic field penetration depth,
whereas the latter one is related to the magnetic field penetration
in the c-axis direction. In this paper we concentrate on the
in-plane electromagnetic response based on the kinetic energy driven
superconductivity and do not consider c-axis properties, which can
be discussed, e.g., by taking into account hopping between adjacent
copper-oxides layers within the tunneling Hamiltonian approach
\cite{sheehy04}.

The paper is organized as follows. Within the framework of the
kinetic energy driven d-wave superconductivity \cite{feng0306}, we
discuss the electromagnetic response of cuprate superconductors in
Section \ref{emressect}, deriving  the kernel of the linear response
with a purely transverse vector potential. In Section
\ref{specular}, based on the specular reflection model
\cite{landau80,tinkham96}, we calculate the temperature and doping
dependence of quantitative characteristics of the electromagnetic
response, such as the local magnetic field profile, the magnetic
field penetration depth, and the superfluid density. Our results
show that the electromagnetic response in cuprate superconductors
can be understood within the framework of the kinetic energy driven
d-wave SC mechanism in the presence of a weak external magnetic
field. We conclude the paper with a brief summary in Section
\ref{conclusions}. In Appendix \ref{invapp} we present a method to
generalize the approach in order to obtain gauge invariant results.

\section{Electromagnetic response in cuprate superconductors}
\label{emressect}

A common feature of cuprate superconductors is the presence of
two-dimensional CuO$_{2}$ planes, and it is believed that the
unconventional physical properties of these systems are closely
related to the doped CuO$_{2}$ plane \cite{damascelli03}. It has
been argued that the essential physics of the doped CuO$_{2}$ plane
\cite{anderson87,damascelli03} is captured by the \emph{t--J} model
on a square lattice. However, for discussions of the electromagnetic
response in cuprate superconductors, the \emph{t--J} model can be
extended by including the exponential Peierls factors as,
\begin{eqnarray}\label{tjham}
H&=&-t\sum_{l\hat{\eta}\sigma}e^{-i({e}/{\hbar})\vv{A}(l)\cdot
\hat{\eta}}C^{\dagger}_{l\sigma} C_{l+\hat{\eta}\sigma}+\mu
\sum_{l\sigma} C^{\dagger}_{l\sigma}C_{l\sigma}\nonumber\\
&+&J\sum_{l\hat{\eta}}{\bf S}_{l}\cdot {\bf S}_{l+\hat{\eta}},
\end{eqnarray}
where $\hat{\eta}=\pm\hat{x},\pm\hat{y}$,  $C^{\dagger}_{l\sigma}$
($C_{l\sigma}$) is the electron creation (annihilation) operator,
${\bf S}_{l}=(S^{x}_{l},S^{y}_{l}, S^{z}_{l})$ are spin operators,
$\mu$ is the chemical potential, and the exponential Peierls factors
account for the coupling of electrons to the weak external magnetic
field in terms of the vector potential $\vv{A}(l)$
\cite{hirsch92,misawa94}. This $t$-$J$ model is subject to an
important local constraint $\sum_{\sigma}
C^{\dagger}_{l\sigma}C_{l\sigma}\leq 1$ in order to avoid the double
occupancy. The strong electron correlation in the $t$-$J$ model
manifests itself by this local constraint \cite{anderson87}, which
can be treated properly in analytical calculations within the
charge-spin separation (CSS) fermion-spin theory \cite{feng0304},
where the constrained electron operators are decoupled as
$C_{l\uparrow}= h^{\dagger}_{l\uparrow}S^{-}_{l}$ and
$C_{l\downarrow}= h^{\dagger}_{l\downarrow} S^{+}_{l}$, with the
spinful fermion operator $h_{l\sigma}=e^{-i\Phi_{l\sigma}}h_{l}$
representing the charge degree of freedom together with some effects
of spin configuration rearrangements due to the presence of the
doped hole itself (charge carrier), while the spin operator $S_{l}$
represents the spin degree of freedom. In particular, it has been
shown that under the decoupling scheme, this CSS fermion-spin
representation is a natural representation of the constrained
electron defined in the Hilbert subspace without double electron
occupancy \cite{feng07}. The advantage of this CSS fermion-spin
approach is that the electron single occupancy local constraint is
satisfied in analytical calculations. Furthermore, these charge
carriers and spins are gauge invariant, and in this sense, they are
real and can be interpreted as the physical excitations
\cite{laughlin97}. In this CSS fermion-spin representation, the
\emph{t--J} model (\ref{tjham}) can be expressed as,
\begin{eqnarray}\label{cssham}
H&=&t\sum_{l\hat{\eta}}e^{-i({e}/{\hbar})\vv{A}(l)\cdot\hat{\eta}}
(h^{\dagger}_{l+\hat{\eta}\uparrow}h_{l\uparrow}S^{+}_{l}
S^{-}_{l+\hat{\eta}}\nonumber\\
&+&h^{\dagger}_{l+\hat{\eta}\downarrow}
h_{l\downarrow}S^{-}_{l}S^{+}_{l+\hat{\eta}}) \nonumber\\
&-&\mu\sum_{l\sigma}h^{\dagger}_{l\sigma}h_{l\sigma}+J_{{\rm eff}}
\sum_{l\hat{\eta}}{\bf S}_{l}\cdot {\bf S}_{l+\hat{\eta}},
\end{eqnarray}
where $J_{{\rm eff}}=(1-\delta)^{2}J$, and $\delta=\langle
h^{\dagger}_{l\sigma}h_{l\sigma}\rangle=\langle h^{\dagger}_{l}
h_{l}\rangle$ is the charge carrier doping concentration. As an
important consequence, the kinetic energy term in the \emph{t--J}
model has been transferred as the charge carrier-spin interaction,
which reflects that even the kinetic energy term in the \emph{t--J}
Hamiltonian has strong Coulomb contribution due to the restriction
of no double occupancy of a given site.

In the case of zero magnetic field, we \cite{feng0306} have shown in
terms of Eliashberg's strong coupling theory \cite{mahan00} that the
charge carrier-spin interaction from the kinetic energy term in the
\emph{t--J} model (\ref{cssham}) induces a charge carrier pairing
state with the d-wave symmetry by exchanging spin excitations in the
higher power of the charge carrier doping concentration $\delta$,
then the SC transition temperature is identical to the charge
carrier pair transition temperature. Moreover, it has been shown
that this SC state is the conventional BCS-like with the d-wave
symmetry \cite{feng07,guo07}, so that the basic BCS formalism with
the d-wave SC gap function is still valid in quantitatively
reproducing all main low energy features of the SC coherence of the
quasiparticle peaks in cuprate superconductors, although the pairing
mechanism is driven by the kinetic energy by exchanging spin
excitations, and other exotic magnetic scattering
\cite{dai01,arai99} is beyond the BCS formalism. Following the
previous discussions \cite{feng0306,feng07,guo07}, the full charge
carrier diagonal and off-diagonal Green's functions in the SC state
can be obtained explicitly in the Nambu representation as,
\begin{eqnarray}
\mathbb{G}({\bf{k}},i\omega_n)=Z_{\rm{hF}}\,\frac{i\omega_n\tau_0 +
\bar{\xi}_{\bf{k}}\tau_3 - \bar{\Delta}_{\rm{hZ}}({\bf{k}})
\tau_1}{(i\omega_n)^2 - E_{{\rm{h}}{\bf{k}}}^2},
\label{holegreenfunction}
\end{eqnarray}
where $\tau_{0}$ is the unit matrix, $\tau_{1}$ and $\tau_{3}$ are
Pauli matrices, the renormalized charge carrier excitation spectrum
$\bar{\xi}_{{\bf k}} =Z_{\rm hF}\xi_{\bf k}$, with the mean-field
(MF) charge carrier excitation spectrum $\xi_{{\bf k}} =Zt\chi
\gamma_{{\bf k}}- \mu$, the spin correlation function $\chi=\langle
S_{i}^{+} S_{i+\hat{\eta}}^{-}\rangle$, $\gamma_{{\bf k}}=
(1/Z)\sum_{\hat{\eta}}e^{i{\bf k}\cdot \hat{\eta}}$, $Z$ is the
number of the nearest neighbor sites, the renormalized charge
carrier d-wave pair gap function $\bar{\Delta}_{\rm hZ}({\bf
k})=Z_{\rm hF} \bar{\Delta}_{\rm h}({\bf k})$, where the effective
charge carrier d-wave pair gap function $\bar{\Delta}_{\rm h}({\bf
k})=\bar{\Delta}_{\rm h} \gamma^{(d)}_{{\bf k}}$ with
$\gamma^{(d)}_{{\bf k}}=({\rm cos} k_{x}-{\rm cos}k_{y})/2$, and the
charge carrier quasiparticle spectrum $E_{{\rm{h}}{\bf k}}=\sqrt
{\bar{\xi}^{2}_{{\bf k}}+ |\bar{\Delta}_{\rm hZ}({\bf k})|^{2}}$,
while the charge carrier quasiparticle coherent weight $Z_{\rm hF}$
and the effective charge carrier gap parameter $\bar{\Delta}_{\rm h}
$ have been determined self-consistently along with another seven
quantities and correlation functions \cite{feng0306,feng07,guo07}.
Let us emphasize that the quasiparticle coherent weight
renormalizing the physical quantities naturally emerges in our
formalism (3), and then both the SC gap function and the
quasiparticle coherence control the SC state. Therefore in our
approach there is no need to introduce any phenomenological charge
renormalization factors in order to describe the electromagnetic
response \cite{sheehy04}.

Now we turn to the discussion of the electromagnetic response in the
kinetic energy driven cuprate superconductors. The weak external
magnetic field applied to the system usually represents a weak
perturbation, but the induced field generated by supercurrents
cancels this weak external field over most of the volume of the
sample. Consequently, the net field acts only very near the surface
on a scale of the magnetic field penetration depth and so it can be
treated as a weak perturbation on the system as a whole. Therefore
the electromagnetic response can be successfully studied within the
linear response approach \cite{fetter71,fukuyama69}, where the
averaged value $\vv{J}$ of the induced microscopic screening current
$\vv{j}$ in the presence of the vector potential $\vv{A}$ is found
as,
\begin{equation}\label{linres}
J_\mu({\bf{q}},\omega)=-\sum\limits_{\nu=1}^3
K_{\mu\nu}({\bf{q}},\omega)A_\nu({\bf{q}},\omega),
\end{equation}
where $K_{\mu\nu}$ is the kernel of the response function and the
Greek indices label the axes of the Cartesian coordinate system.
Recall that, as always in the linear response method, the thermal
average of the supercurrent is calculated with the unperturbed
Hamiltonian, i.e. for $\vv{A}\equiv0$ in Eq. (\ref{cssham}). Let us
also note that the relation (\ref{linres}), which is local in the
reciprocal space, in general implies a nonlocal response in the
coordinate space.

The kernel, which plays a central role in the description of the
electromagnetic response, and once known allows us to calculate
quantitative characteristics of the electromagnetic response, can be
separated into two parts:
\begin{equation}\label{kernel}
K_{\mu\nu}({\bf{q}},\omega) = K^{({\rm{d}})}_{\mu\nu}({\bf{q}},
\omega) + K^{({\rm{p}})}_{\mu\nu}({\bf{q}},\omega),
\end{equation}
a diamagnetic part $K^{({\rm{d}})}_{\mu\nu}$ and a paramagnetic one
$K^{({\rm{p}})}_{\mu\nu}$. The evaluation of the diamagnetic
contribution usually poses no difficulties since it is known almost
immediately from the form of the diamagnetic current operator: it
turns out to be diagonal and proportional to the average kinetic
term.  However, the paramagnetic part can only be calculated
approximately since it involves evaluation of a retarded
current-current correlation function (polarization bubble). As the
retarded function is inconvenient for perturbation analysis one
usually proceeds with the corresponding imaginary-time-ordered
Matsubara function,
\begin{equation}\label{corP}
P_{\mu\nu}(\vv{q},\tau)=-\langle T_\tau
\{j^{({\rm{p}})}_{\mu}(\vv{q},\tau) j_{\nu}^{({\rm{p}})}(-\vv{q},0)
\}\rangle ,
\end{equation}
where the paramagnetic current operator is defined in the imaginary
time $\tau$ Heisenberg picture. Hence, the main problem is reduced
to the evaluation of a retarded current commutator for the
unperturbed system. The retarded current-current correlation
function is then obtained in a standard way from the imaginary time
Fourier transform $P_{\mu\nu}(\vv{q},i\omega_n)$ of the Matsubara
function (\ref{corP}) by analytic continuation to real frequencies
\cite{mahan00}.

The vector potential $\vv{A}$ (then the weak external magnetic field
$B=rot\vv{A}$) has been coupled to the electrons, which are now
represented by $C_{l\uparrow}= h^{\dagger}_{l\uparrow}S^{-}_{l}$ and
$C_{l\downarrow}= h^{\dagger}_{l\downarrow}S^{+}_{l}$ in the CSS
fermion-spin representation. However, in the CSS framework, the
vector potential $\vv{A}$ is coupled to $h^{\dagger}_{l\sigma}$,
while the corresponding weak external magnetic field ${\bf B}
=rot\vv{A}$ is coupled to ${\bf S}_{l}$ by including the Zeeman term
\cite{zhang09} in the Hamiltonian (\ref{tjham}). For cuprate
superconductors, the upper critical magnetic field is 50 Tesla or
greater around the optimal doping. In this paper, we mainly focus on
the case where the applied external magnetic field $B<10$ mT is much
less than the upper critical magnetic field. In this case, the
Zeeman term \cite{zhang09} in the Hamiltonian (\ref{tjham}) has been
dropped, and then the electron current operator
$j_\mu=j_\mu^{(\rm{d})}+j_\mu^{(\rm{p})}$ can be obtained by
differentiating the Hamiltonian (\ref{cssham}) with respect to the
vector potential as,
\begin{subequations}
\begin{eqnarray}
j_\mu^{(\rm{d})}&=&\frac{\chi e^2 t}{2\hbar^2}\sum\limits_{l\sigma}
\left( \osd{h}{l+\hat{\mu}}\os{h}{l}+\osd{h}{l}\os{h}{l+\hat{\mu}}
\right)A_\mu(l), ~~~~\label{tcurdia}\\
j_\mu^{(\rm{p})}&=&-\frac{i\chi e t}{2\hbar} \sum\limits_{l\sigma}
\left(\osd{h}{l+\hat{\mu}}\os{h}{l} - \osd{h}{l}\os{h}{l+\hat{\mu}}
\right),\label{tcurpara}
\end{eqnarray}
\end{subequations}
being the diamagnetic and paramagnetic contributions, respectively.

Since the diamagnetic current is explicitly proportional to the
vector potential, it is straightforward to find the diamagnetic part
of the response kernel as,
\begin{eqnarray}\label{diakernel}
K_{\mu\nu}^{(\rm{d})}(\vv{q})&=&-\frac{Z_{\rm{hF}}\chi e^2t}
{\hbar^2}{1\over N} \sum \limits_{\vv{k}} \delta_{\mu\nu}\cos k_\mu
\nonumber\\
&\times&\left(1-\frac{\bar{\xi}_{\bf{k}}}{E_{{\rm{h}}\bf{k}}}
\tanh{\frac{\beta E_{{\rm{h}}\bf{k}}}{2}} \right)
\nonumber\\
&=&-\frac{2\chi\phi e^2t}{\hbar^2}\,\delta_{\mu\nu}.~~~~~~
\end{eqnarray}

The paramagnetic part of the response kernel is more complicated to
calculate, as it involves evaluation of the current-current
correlation function (\ref{corP}). In particular, if we want to keep
the theory gauge invariant, it is crucial to approximate the
correlation function in a way maintaining local charge conservation
\cite{fukuyama69,schrieffer83,misawa94,arseev06}. Since in the
following calculations we will work with a fixed gauge of the vector
potential, we postpone the detailed discussion of this problem until
Appendix \ref{invapp}. Starting with the paramagnetic current
operator (\ref{tcurpara}), we can rewrite its Fourier transform in
the notation of Nambu fields $ \Psi^\dagger_\vv{k} =
\left(\oud{h}{\vv{k}}, \od{h}{-\vv{k}}\right) $ and $ \Psi_{\vv{k} +
\vv{q}} = \left( \ou{h}{\vv{k}+\vv{q}}, \odd{h}{-\vv{k}-\vv{q}}
\right)^T $ as
\begin{equation}\label{curnambu}
j^{(\rm{p})}_\mu(\vv{q})={1\over N}\sum\limits_{\vv{k}}
\Psi_{\vv{k}}^\dagger \left[-\frac{\chi et}{\hbar}\,
e^{i\frac{q_\mu}{2}}\sin\left(k_\mu +
\frac{q_\mu}{2}\right)\tau_0\right] \Psi_{\vv{k}+\vv{q}}.
\end{equation}
For the purpose of the discussion addressing the gauge invariance
problem, presented in Appendix \ref{invapp}, it is convenient to
find the charge density in the Nambu notation as well. Within the
CSS fermion-spin scheme, we first find $ \rho(\vv{q})\approx
-({e}/{2N}) \sum_{\vv{k}} (\delta_{\vv{q},0}-\oud{h}{\vv{k}}
\ou{h}{\vv{k}+ \vv{q}}-\odd{h}{\vv{k}}\od{h}{\vv{k}+\vv{q}})$. Then
the paramagnetic four-current operator can be represented in the
Nambu form as $ j_\mu^{\rm{(p)}}(\vv{q}) = \sum\limits_{\vv{k}}
\Psi_{\vv{k}}^\dagger \gamma_\mu(\vv{k+q},\vv{k})
\Psi_{\vv{k}+\vv{q}}$, where the bare current vertex,
\begin{equation}
{\mathbf{\gamma}}_\mu(\vv{k}+\vv{q},\vv{k})= \left
\{\begin{array}{ll} -\frac{\chi et}{\hbar}\,
e^{i\frac{q_\mu}{2}}\sin\left(k_\mu + \frac{q_\mu}{2}\right)\tau_0
                            & {\rm{for}}\ \mu\neq0\\
-\frac{e}{2}\, \tau_3      & {\rm{for}}\ \mu=0.\\
\end{array}\right.\label{barevertex}
\end{equation}
It is necessary to be aware that we are calculating the polarization
bubble with the paramagnetic current operator (\ref{curnambu}),
i.e., bare current vertices (\ref{barevertex}), but charge carrier
Green functions. Consequently, as in this scenario we do not take
into account longitudinal excitations properly
\cite{schrieffer83,misawa94}, the obtained results are valid only in
the gauge, where the vector potential is purely transverse, e.g. in
the Coulomb gauge. In this case, we can obtain the correlation
function (\ref{corP}) in the Matsubara representation as,
\begin{widetext}
\begin{eqnarray}
P_{\mu\nu}(\vv{q},i\omega_n) &=& \left( \frac{\chi e
t}{\hbar}\right)^2 e^{\frac{i}{2}(q_\mu -q_\nu)}{1\over N}
\sum\limits_{\vv{k}} \sin\left(k_\mu +
\frac{q_\mu}{2}\right)\sin\left(k_\nu + \frac{q_\nu}{2}\right)
\frac{1}{\beta}\sum\limits_{i\nu_m}
{\rm{Tr}}\,\left[{\mathbb{G}}(\vv{k+q},i\omega_n+i\nu_m)
{\mathbb{G}}(\vv{k},i\nu_m) \right].~~~~\label{barepolmats}
\end{eqnarray}
\end{widetext}
Restricting the discussion to the static limit ($\omega\sim 0$) and
completing the summation over Matsubara frequencies, we obtain the
bare vertex current-current correlation function, and hence the
paramagnetic part of the response kernel as,
\begin{widetext}
\begin{eqnarray}
K_{\mu\nu}^{(\rm{p})}(\vv{q},0) & = & -\left(\frac{\chi et
Z_{\rm{hF}}} {\hbar^2}\right)^2e^{\frac{i}{2}(q_\mu - q_\nu)}
{1\over N}\sum\limits_{\vv{k}} \sin\left(k_\mu +
\frac{q_\mu}{2}\right)\sin\left(k_\nu +
\frac{q_\nu}{2}\right)\nonumber\\
&\times&\left\{\frac{1}{\Ek + \Ekq}\left[ 1 - \frac{\xikq\xik +
\Dkq\Dk}{\Ek\Ekq} \right]\left[1- \nF(\Ek) - \nF(\Ekq)
\right]\right. \nonumber\\
&+& \left.\frac{1}{\Ek - \Ekq}\left[ 1 + \frac{\xikq\xik +
\Dkq\Dk}{\Ek\Ekq} \right] \left[\nF(\Ekq) - \nF(\Ek) \right]
\right\}.~~~~~ \label{parakernel}
\end{eqnarray}
\end{widetext}
Note that in the long wavelength limit, when $|\vv{q}|\to0$, the
former term in Eq. (\ref{parakernel}) vanishes, and the latter turns
into $ -2\left({\chi et Z_{\rm{hF}}}/{N\hbar^2}\right)^2
\sum\limits_{\vv{k}} \sin k_\mu \sin k_\nu\, \nF(\Ek)[1-\nF(\Ek)]$,
which is equal to zero in the zero-temperature limit. Hence, in this
case, the long wavelength electromagnetic response at zero
temperature is determined by the diamagnetic part of the kernel
only.

\section{Quantitative characteristics}
\label{specular}

\begin{figure}[h!]
\includegraphics[scale=0.5]{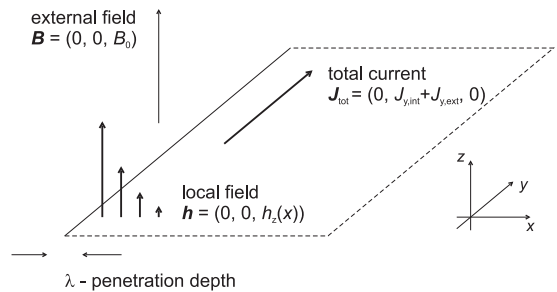}
\caption{Geometry of the specular reflection model. The current
$\vv{J}_{\rm{ext}}$ simulates external magnetic field at the edge of
the sample ($x=0$), whereas the induced supercurrent
$\vv{J}_{\rm{int}}$ is the (linear) reaction of the
system.\label{specularfig}}
\end{figure}

The way the system reacts to a weak electromagnetic stimulus is
entirely described by the linear response kernel, which is
calculated within a microscopic model. Once the kernel is known, the
effect of a weak external magnetic field can be quantitatively
characterized by experimentally measurable quantities such as the
magnetic field penetration depth and the local magnetic field
profile. Technically, we need to combine one of the Maxwell
equations with the relation (\ref{linres}) describing the response
of the system and solve them together for the vector potential. This
is the step in which a particular gauge of the vector
potential---usually implied by the geometry of the system---is set.
However, the kernel function derived within the linear response
theory describes the response of an \emph{infinite} system.  In
order to take into account the confined geometry of cuprate
superconductors it is necessary to introduce a surface being the
boundary between the environment and the sample. This can be done
within the standard specular reflection model
\cite{landau80,tinkham96} with a two-dimensional geometry of the SC
plane, in the configuration with external magnetic field
perpendicular to the ab plane, as shown in Fig. \ref{specularfig}.
In the present paper we study magnetic field penetration effects
within the ab plane only, so our primary goal is to find and discuss
the magnetic field in-plane penetration depth.

\begin{figure}[h!]
\includegraphics[scale=0.75]{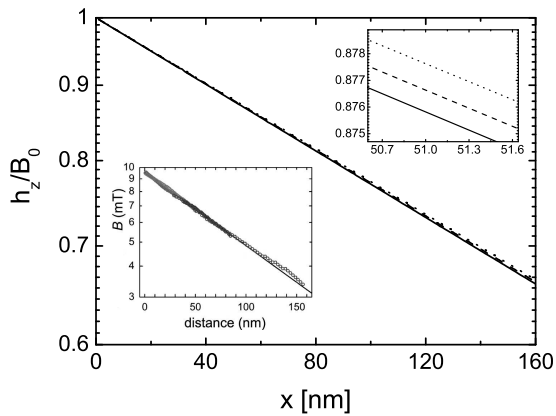}
\caption{The local magnetic field profile as a function of the
distance from the surface at temperature $T=0.02 J$ for doping
concentration $\delta=0.150$ (solid line), $\delta=0.147$ (dashed
line), and $\delta=0.144$ (dotted line) with parameter $t=2.5J$.
Inset (top): zoom into the intermediate range of the local magnetic
field profile. Inset (bottom): the corresponding experimental result
for YBa$_2$Cu$_3$O$_{7-y}$ taken from Ref. \onlinecite{suter04}.
\label{profilefig}}
\end{figure}

In order to simulate an external magnetic field at the surface of a
two-dimensional sample, we introduce an external current sheet
$J_{y,\rm{ext}}(x)=-2B_0\delta(x)/\mu_0$ at the edge $x=0$, where
$\mu_0$ is the magnetic permeability and $B_0$ is the amplitude of
the weak external magnetic field at the surface ($x=0$). From the
Maxwell equation for the curl of the local magnetic field
${\rm{rot}}\,{\vv{h}}=\mu_0 (\vv{J}_{\rm{int}}+\vv{J}_{\rm{ext}})=
\mu_0 \vv{J}_{\rm{int}}+[0,-2B_0\delta(x),0]$ and the fact, that the
induced supercurrent $\vv{J}_{\rm{int}}$ flows along the $y$ axis,
we can state that the local magnetic field is of the form
$\vv{h}(\vv{r})=[0,0,h_z(x)]$. In order to discuss the magnetic
field penetration effect, spatial dependence of the local magnetic
field has to be found. Let us begin with the identity $ \rm{rot}\,
\rm{rot}\,\vv{A}={\rm{grad}\,\rm{div}\,\vv{A}}- \nabla^2 \vv{A} $
and choose the vector potential as $\vv{A}(\vv{r})= [0,A_{y}(x),0]$
setting the Coulomb gauge. In this case, $ q_x^2 A_y(\vv{q}) =
\mu_0\left[ J_{y,\rm{int}}(\vv{q}) + J_{y,\rm{ext}}(\vv{q})\right],
$ because the vector potential has only non-zero $y$ component.
Finally, including the form of the external current, the linear
relation (\ref{linres}) between the induced supercurrent and the
vector potential $J_{y,\rm{int}}(\vv{q}) =
-K_{yy}(\vv{q})A_y(\vv{q})$, and solving for the vector potential we
obtain,
\begin{equation}\label{aspec}
A_y(\vv{q})=-8\pi^2B_0\,\frac{\delta(q_y)\delta(q_z)}{\mu_0
K_{yy}(\vv{q}) + q_x^2}.
\end{equation}
Since the vector potential has only the $y$ component, the only
non-zero component of the local magnetic field $\vv{h} =
\rm{rot}\,\vv{A}$ is that along the $z$ axis and $h_z(\vv{q}) =
iq_xA_y(\vv{q})$. Substituting the derived form of the vector
potential (\ref{aspec}), and taking the inverse Fourier transform,
the local magnetic field profile can be obtained as,
\begin{equation}\label{profile}
h_z(x) =  \frac{B_0}{\pi}\int\limits_{-\infty}^\infty {\rm{d}}q_x\,
\frac{q_x \sin q_xx}{\mu_0 K_{yy}(q_x,0,0) + q_x^2}.
\end{equation}
Local magnetic field profiles can be measured experimentally, e.g.
using the muon-spin rotation technique \cite{khasanov04,suter04},
providing an important tool to investigate the details of magnetic
field screening inside the sample. In cuprate superconductors the
screening is found to be of exponential character
\cite{khasanov04,suter04}, in support of a local (London-type)
nature of the electrodynamics \cite{schrieffer83}. For the
convenience of the following discussions, we introduce a
characteristic length scale $a_0 =\sqrt{\hbar^2 a/\mu_0 e^2 J}$.
Using a reasonably estimative value of $J/k_{\rm{B}}\approx 1000$K
and $a\approx 0.383$nm, which is the lattice parameter for the
cuprate superconductor YBa$_2$Cu$_3$O$_{7-y}$, we obtain $a_0
\approx 97.8$nm. In Fig. \ref{profilefig}, we plot the local
magnetic field profile (\ref{profile}) as a function of the distance
from the surface at temperature $T=0.02J$ for the doping
concentration $\delta=0.150$ (solid line), $\delta=0.147$ (dashed
line), and $\delta=0.144$ (dotted line) with parameter $t=2.5J$. For
comparison, the corresponding experimental result \cite{suter04} of
the local magnetic field profiles for the high quality
YBa$_2$Cu$_3$O$_{7-y}$ sample is also shown in Fig. \ref{profilefig}
(inset, bottom). If a weak external field $B_0\approx 10$ mT is
applied to the system just as it has been done in the experimental
measurement \cite{suter04}, then the experimental result
\cite{suter04} for YBa$_2$Cu$_3$O$_{7-y}$ is well reproduced. In
particular, our theoretical results perfectly follow an exponential
law as expected for the local electrodynamic response.

\begin{figure}[h!]
\includegraphics[scale=0.75]{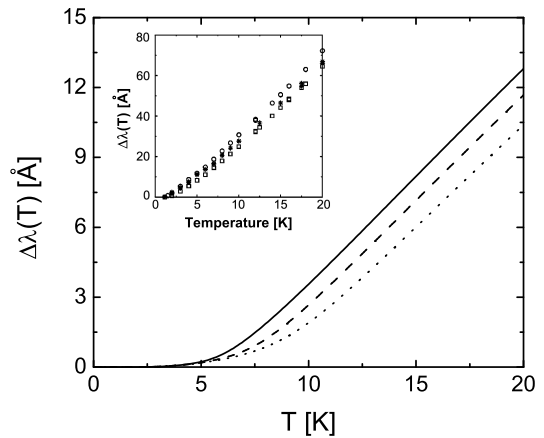}
\caption{Temperature dependence of the magnetic field in-plane
penetration depth $\Delta\lambda(T)$ for the doping concentration
$\delta=0.150$ (solid line), $\delta=0.149$ (dashed line), and
$\delta=0.148$ (dotted line) with parameter $t/J=2.5$. Inset: the
corresponding experimental data for YBa$_2$Cu$_3$O$_{7-y}$ taken
from Ref. \onlinecite{kamal98}. \label{lambdafig}}
\end{figure}

The above obtained local magnetic field profile $h_z(x)$ allows us
to determine the magnetic field in-plane penetration depth
$\lambda(T)$ in a straightforward way. According to the definition
$\lambda(T)=B_0^{-1} \int_0^\infty h_z(x)\,{\rm{d}}x$, the magnetic
field in-plane penetration depth can be evaluated as,
\begin{equation}\label{lambda}
\lambda(T) = \frac{2}{\pi} \int\limits_0^\infty
\frac{{\rm{d}}q_x}{\mu_0 K_{yy}(q_x,0,0) + q_x^2}.
\end{equation}
In this case, we obtain the zero-temperature magnetic field in-plane
penetration depth $\lambda(0)\approx 380.8$nm for the doping
concentration $\delta=0.150$ with parameter $t/J=2.5$. This
anticipated value is very close to the values of the magnetic field
in-plane penetration depth $\lambda\approx 156$nm $\sim 400$nm
observed in different families of cuprate superconductors
\cite{bernhard01,khasanov04,uemura93}. Furthermore,
$\Delta\lambda(T)=\lambda(T)-\lambda(0)$ as a function of
temperature $T$ for the doping concentration $\delta=0.150$ (solid
line), $\delta=0.149$ (dashed line), and $\delta=0.148$ (dotted
line) with parameter $t/J=2.5$ is plotted in Fig. \ref{lambdafig} in
comparison with the corresponding experimental results
\cite{kamal98} of YBa$_2$Cu$_3$O$_{7-y}$ (inset). Our theoretical
results show linear characteristics of the magnetic field in-plane
penetration depth $\Delta\lambda(T)$, except for extremely low
temperatures where a strong deviation from the linear
characteristics (a nonlinear effect) appears. In particular, this
crossover from the linear temperature dependence in the low
temperature regime into the nonlinear one at extremely low
temperatures is observed experimentally in nominally clean crystals
of cuprate superconductors
\cite{bonn96,kamal98,jackson00,panagopoulos99,pereg07,khasanov04,suter04,sonier99}.
Apparently, there is a substantial difference between theory and
experiment, namely, the value of the difference between $\lambda(T)$
and $\lambda(0)$ calculated theoretically is much smaller than the
corresponding value measured in the experiment. However, upon a
closer examination one can see immediately that the main difference
is due to fact that the calculated $\lambda(T)$ increases slowly
with temperature. As for a qualitative discussion in this paper, the
overall tendency seen in the theoretical result is consistent with
that in the experiment \cite{kamal98}. In cuprate superconductors,
the values of $J$ and $t$ are believed to vary somewhat from
compound to compound \cite{damascelli03}. Therefore the quantitative
agreement can be reached by adjustments of theory's parameters $t$
and $J$, or by introducing the next neighbor hopping $t'$.

\begin{figure}[h!]
\includegraphics[scale=0.75]{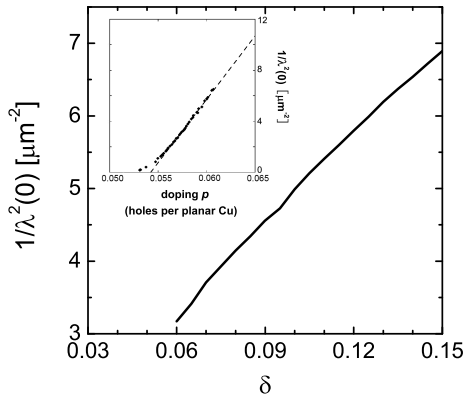}
\caption{Doping dependence of the zero-temperature in-plane
superfluid density in the underdoped regime with $t/J=2.5$. Inset:
the corresponding experimental result for YBa$_2$Cu$_3$O$_{7-y}$
taken from Ref. \onlinecite{broun07}. \label{rhofig}}
\end{figure}

A quantity which is closely related to the magnetic field in-plane
penetration depth $\lambda(T)$ is the in-plane superfluid density
$\rho_{\rm s}(T)\equiv \lambda^{-2}(T)$. For a better understanding
of the physical properties of cuprate superconductors, we have
calculated the doping dependence of the zero-temperature in-plane
superfluid density $\rho_{\rm s}(0)$ in the underdoped regime. The
result for parameter $t/J=2.5$ is plotted in Fig. \ref{rhofig} in
comparison with the corresponding experimental data \cite{broun07}
for YBa$_2$Cu$_3$O$_{7-y}$ (inset). It is shown that the in-plane
superfluid density $\rho_{\rm s}(0)$ in the underdoped regime
vanishes more or less linearly with decreasing doping concentration
$\delta$, in qualitative agreement with experimental results of
cuprate superconductors \cite{uemura8991,broun07,bernhard01}. This
result also is a natural consequence of the linear doping dependence
of the SC transition temperature $T_{c}\propto \delta$ in the
underdoped regime in the framework of the kinetic energy driven SC
mechanism \cite{feng0306}, where the SC transition temperature
$T_{c}$ is set by the charge carrier doping concentration, and then
the density of the charge carriers directly determines the in-plane
superfluid density in the underdoped regime.

The appearance of the nonlinearity in the temperature dependence of
the magnetic field in-plane penetration depth in cuprate
superconductors at extremely low temperatures, as shown in Fig.
\ref{lambdafig}, can be attributed to the nonlocal effects, which in
the case of a pure d-wave cuprate superconductor with nodes in the
gap become significant for the electromagnetic response
\cite{yip92,kosztin97,franz97,li00,sheehy04}. In general, the
relation between the supercurrent and the vector potential
(\ref{linres}) is nonlocal in the coordinate space due to the finite
size of charge carrier Cooper pairs. In the framework of the kinetic
energy driven d-wave SC mechanism, the size of charge carrier pairs
in the clean limit is of the order of the coherence length
$\zeta(\vv{k})=\hbar v_{\rm F}/\pi \Delta_{\rm h}(\vv{k})$, where
$v_{\rm{F}}=\hbar^{-1}\partial\xi_{\bf k}/\partial {\bf k}|_{k_{F}}$
is the charge carrier velocity at the Fermi surface, and therefore
the size of charge carrier pairs is momentum dependent. Although the
weak external magnetic field decays exponentially on the scale of
the magnetic field in-plane penetration length $\lambda(T)$, any
nonlocal contributions to measurable quantities are of the order of
$\kappa^{-2}$, where $\kappa$, known as the Ginzburg--Landau
parameter, is the ratio of the magnetic field in-plane penetration
depth $\lambda$ and the coherence length $\zeta$. However, in the
d-wave cuprate superconductors, the characteristic feature is the
existence of four nodal points $[\pm\pi/2,\pm\pi/2]$ in the
Brillouin zone, where the charge carrier gap function vanishes
$\Delta_{\rm h}(\vv{k}) |_{[\pm\pi/2,\pm\pi/2]}=\Delta_{\rm h}({\rm
cos}k_{x}-{\rm cos}k_{y})/2|_{[\pm\pi/2,\pm\pi/2]}=0$. As a
consequence, the coherence length $\zeta(\vv{k})$ diverges around
the nodes. In particular, at extremely low temperatures, the
quasiparticles selectively populate the nodal region, and the major
contribution to measurable quantities comes from these
quasiparticles. In this case, the Ginzburg--Landau ratio
$\kappa(\vv{k})$ around the nodal region is no longer large enough
for the system to belong to the class of type-II superconductors,
and the condition of the local limit is not fulfilled
\cite{kosztin97}. On contrary, the system falls then into the
extreme nonlocal limit, and therefore the nonlinear characteristic
in the temperature dependence of the magnetic field in-plane
penetration depth can be observed experimentally in cuprate
superconductors at sufficiently low temperatures
\cite{bonn96,khasanov04,suter04,sonier99}. However, with increasing
temperature, the quasiparticles around the nodal region become
excited out of the condensate, and the nonlocal effect fades away.
In this case, the momentum dependent coherence length
$\zeta(\vv{k})$ can be replaced approximately with the isotropic one
$\zeta_0=\hbar v_{\rm F}/ \pi \Delta_{\rm h}$. Then the
Ginzburg--Landau parameter $\kappa_0\approx \lambda(0)/\zeta_0
\approx 180$, and the condition for the local limit is satisfied.
This anticipated value of the Ginzburg--Landau parameter $\kappa_0
\approx 180$ is not too far from the range $\kappa_0 \approx 150\sim
400$ estimated experimentally for different families of cuprate
superconductors \cite{bernhard01,khasanov04,uemura93}. Consequently,
the cuprate superconductors at moderately low temperatures turn out
to be type-II superconductors, where nonlocal effects are
negligible, the electrodynamics is purely local and the magnetic
field decays exponentially over a length of the order of a few
hundreds nm. In this local limit, the pure d-wave pairing state in
the kinetic energy driven SC mechanism gives the magnetic field
penetration depth $\lambda(T)\propto T$ \cite{tsuei00,kosztin97}.
This is why the linear temperature dependence of the magnetic field
in-plane penetration depth $\lambda(T)$ is observed experimentally
\cite{bonn96,hardy93,kamal98,jackson00,panagopoulos99,pereg07} in
cuprate superconductors at moderately low temperatures.

Finally, we have to note that a deviation from the linear Uemura
relation between the in-plane superfluid density $\rho_{\rm s}(0)$
and doping concentration $\delta$ has been observed recently in the
underdoped regime \cite{broun07,pereg07,hardy04}. This deviation
from the linear Uemura relation suggests a sublinear dependence of
the critical temperature $T_{\rm{c}}$ and the superfluid density
$\rho_{\rm s}(0)$, since $T_{\rm{c}}$ must fall to zero when
$\rho_{\rm s}(0)$ does \cite{pereg07,hardy04}. The parent compound
of doped cuprate superconductors is a Mott insulator with an
antiferromagnetic long-range order and superconductivity occurs when
the antiferromagnetic long-range order state is suppressed by doped
charge carriers. Since these doped charge carriers in cuprate
superconductors are induced by the replacement of some ions by other
ones with different valences, or the addition of excess oxygens in
the block layer, therefore, in principle, all cuprate
superconductors have natural impurities \cite{damascelli03}.
Therefore the impurities play an important role in the
electromagnetic response and lead to some subtle differences in the
electromagnetic response for different families of cuprate
superconductors \cite{bonn96}. In this case, the impurity effect on
the SC state of cuprate superconductors is also a possible source
for the deviation from the linear Uemura relation. In this context
we \cite{wang08} have discussed the effect of the extended impurity
scatterers on the quasiparticle transport of cuprate superconductors
in the SC state based on the nodal approximation of the
quasiparticle excitations and scattering processes, and predicted
that in contrast with the dome shape of the doping dependent SC gap
parameter, the minimum of the microwave conductivity occurs around
the optimal doping, and then increases in both underdoped and
overdoped regimes. However, in this paper we are primarily
interested in exploring the general notion of the electromagnetic
response in cuprate superconductors in the SC state. The qualitative
agreement between the present theoretical results in the clean limit
and experimental data for different families of cuprate
superconductors provides an important confirmation of the nature of
the SC phase of cuprate superconductors as a d-wave BCS-like SC
state within the kinetic energy driven SC mechanism.

\section{Conclusions}
\label{conclusions}

In this paper we have discussed the electromagnetic response in
cuprate superconductors within the framework of kinetic energy
driven d-wave superconductivity. Following the linear response
theory and taking into account the two-dimensional geometry of
cuprate superconductors within the specular reflection model, we
have reproduced some main features of the electromagnetic response
experiments on cuprate superconductors, including the exponential
local magnetic field profile, the linear temperature dependence of
the in-plane penetration depth in the low temperature range and its
nonlinear temperature dependence at extremely low temperatures.
Moreover, the linear doping dependence of the zero-temperature
in-plane superfluid density in the underdoped regime has been
reproduced. In particular, we have clearly identified the
limitations of the used approximations, especially with respect to
the problem of gauge invariance. Furthermore, we have proposed a
method to generalize the discussions in order to make them
independent of a particular choice of the vector potential.

\acknowledgments

The authors would like to thank Dr. Zhi Wang and Dr. Yu Lan for
helpful discussions. This work was supported by the National Natural
Science Foundation of China under Grant No. 10774015, and the funds
from the Ministry of Science and Technology of China under Grant
Nos. 2006CB601002 and 2006CB921300. MK gratefully acknowledges
support from a research scholarship funded by Institute of Physics,
Wroc{\l}aw University of Technology.

\appendix
\section{Gauge-invariant electromagnetic response}
\label{invapp}

It is well known that gauge invariance is a direct consequence of
local charge conservation \cite{fukuyama69,schrieffer83}, which is
mathematically expressed by the charge density-current continuity
equation or its Green function analogue called the generalized Ward
identity (GWI) \cite{fukuyama69,schrieffer83,misawa94,arseev06}
\begin{equation}\label{GWI}
-2N \sum\limits_{\mu=0}^3 q_\mu\Gamma_\mu (k+q, k) =
\tau_3\mathbb{G}^{-1}(k) - \mathbb{G}^{-1}(k+q)\tau_3.
\end{equation}
Here $\Gamma_\mu$ is a dressed version of the four-current vertex
function, and the four-vector notation $q=(\vv{q},q_0=i\omega)$
along with the metric $(1,1,1,-1)$ has been introduced.

\begin{figure}[h!]
\includegraphics[scale=0.6]{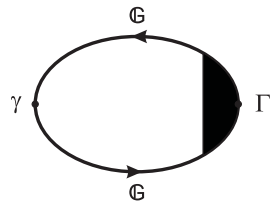}
\caption{Dressed polarization bubble (Nambu notation). Here both the
Green function and the current vertex are dressed with the pairing
interaction due to the spin bubble.\label{drbubblefig}}
\end{figure}

Since the local charge conservation requirement is quite universal
and fundamental, it should be inherent to any theory of the
electromagnetic response which is expected to be gauge invariant.
The purpose of this appendix is to propose---within the framework of
the kinetic energy driven superconductivity---a method to dress the
current vertex in a way, which does not violate the GWI. Once such a
method is found, the bare polarization bubble (\ref{corP}) can be
replaced with its dressed version presented in Fig.
\ref{drbubblefig}, and the resulting kernel of the response function
will provide correct results for any gauge of the vector potential.

In the first step we will note that
\begin{equation}\label{wardmean}
-2N \sum\limits_{\mu=0}^3 q_\mu\gamma_\mu \left(k+q, k\right) =
\tau_3\mathbb{G}^{(0)-1}(k) - \mathbb{G}^{(0)-1}(k+q)\tau_3,
\end{equation}
i.e. that the GWI for the bare current vertex is satisfied with the
MF charge carrier Green function
$\mathbb{G}^{(0)}(k)=[(i\omega_n)^2-\xi_\vv{k}^2]^{-1}
(i\omega_n\tau_0+\xi_\vv{k}\tau_3)$. Substituting the MF charge
carrier Green function, the rhs of Eq. (\ref{wardmean}) turns into
$\tau_3\mathbb{G}^{(0)-1}(k) - \mathbb{G}^{(0)-1}(k+q)\tau_3 =
\left(\xi_\vv{k+q} - \xi_{\vv{k}} \right)\tau_0 - q_0\tau_3$.
Moreover, in the long wavelength limit, after including the explicit
form of the MF charge carrier dispersion relation found within the
framework of the kinetic energy driven superconductivity
\cite{feng07}, it further simplifies to $\tau_3\mathbb{G}^{(0)-1}(k)
-\mathbb{G}^{(0)-1}(k+q) \tau_3\approx \left[-2t\chi(q_x\sin k_x +
q_y\sin k_y)\right] \tau_0 - q_0\tau_3. $ Now, recalling the form of
the bare vertex (\ref{barevertex}), we can notice that in the long
wavelength limit the scalar product on the left-hand side of Eq.
(\ref{wardmean}) $-q_0\gamma_0 + \vv{q}\vv{\gamma} = (2N)^{-1}
\left( \tau_3 q_0 - \tau_0\nabla_\vv{k} \xi_\vv{k}\cdot\vv{q}
\right)$, which proves the equality (\ref{wardmean}).

\begin{figure}[h!]
\includegraphics[scale=0.35]{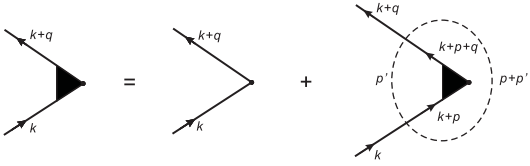}
\caption{Ladder-type approximation for the dressed vertex.
\label{ladderfig}}
\end{figure}

It is well known that in order to obtain a dressed vertex function,
which does not violate the GWI, a ladder-type approximation can be
adapted \cite{schrieffer83,fukuyama69,misawa94}. The nature of the
pairing mechanism \cite{feng0306,feng07}, which originates from the
spin bubble, suggests a ladder-like approximation of the form,
\begin{widetext}
\begin{eqnarray}
\Gamma_\mu(k+q,k)&=& \gamma_\mu(k+q,k) +
\frac{1}{N}\,\frac{1}{\beta} \sum\limits_{p}\tau_3
\mathbb{G}(k+p+q)\Gamma_\mu(k+p+q,k+p){\mathbb{G}}(k+p)\tau_3
\nonumber\\
&\times& \frac{1}{N} \sum\limits_{\vv{p}'}
\Lambda^2_{\vv{p}+\vv{p}'+\vv{k}} \Pi(\vv{p},\vv{p}';ip_m),
~~~~~\label{ladder}
\end{eqnarray}
\end{widetext}
which is graphically presented in Fig. \ref{ladderfig}.

In order to prove that the approximation (\ref{ladder}) for the
dressed vertex in fact implies a gauge invariant description of the
electromagnetic response, it is necessary and sufficient to check
whether it does not violate the GWI (\ref{GWI}). In order to prove
it, we insert the dressed vertex function (\ref{ladder}) into the
left-hand side of Eq. (\ref{GWI}) and use the identity
$-2N\sum_{\mu=0}^3 q_\mu\Gamma_\mu(s+q,s) = \tau_3
{\mathbb{G}}^{-1}(s) - {\mathbb{G}}^{-1}(s+q)\tau_3$ to obtain
\begin{widetext}
\begin{eqnarray}
\sum\limits_{\mu=0}^3 q_\mu \Gamma_\mu(k+q,k) &=&
\sum\limits_{\mu=0}^3 q_\mu \gamma_\mu(k+q,k) +
\frac{1}{N}\,\frac{1}{\beta} \sum\limits_{p} \left(
-\frac{1}{2N}\right) \left[ \tau_3 {\mathbb{G}}(k+p+q) \tau_3 -
\tau_3 {\mathbb{G}}(k+p)\tau_3 \right]\nonumber\\
&\times& \frac{1}{N} \sum\limits_{\vv{p}'}
\Lambda^2_{\vv{p}+\vv{p}'+\vv{k}} \Pi(\vv{p},\vv{p}';ip_m).~~~~~
\label{inveqn}
\end{eqnarray}
\end{widetext}
In the long wavelength limit  we use the approximation
$\Lambda^2_{\vv{p}+\vv{p}'+\vv{k}}\approx
\Lambda^2_{\vv{p}+\vv{p}'+\vv{k}+\vv{q}}$. Then we can simplify Eq.
(\ref{inveqn}) into
$
\sum_{\mu=0}^3 q_\mu \Gamma_\mu(k+q,k) \approx
\sum_{\mu=0}^3 q_\mu \gamma_\mu(k+q,k) 
-\left[ \Sigma(k+q)\tau_3 - \tau_3\Sigma(k)\right]/2N. $ Using the
fact that the free vertex satisfies the GWI with the MF Green
function, as stated in Eq. (\ref{wardmean}), and arranging the terms
with respect to the Pauli matrices, we have
\begin{eqnarray*}
-2N\sum\limits_{\mu=0}^3 q_\mu \Gamma_\mu(k+q,k) &\approx& \tau_3
[\mathbb{G}^{(0)-1}(k)-\Sigma(k)]\\
&-&[\mathbb{G}^{(0)-1}(k+q)- \Sigma(k+q)]\tau_3.
\end{eqnarray*}
Hence, identifying the terms in the square brackets as dressed
charge carrier Green functions, we eventually obtain the GWI
(\ref{GWI}), what proves that the ladder-type approximation
(\ref{ladder}) for the vertex function in the dressed polarization
bubble in Fig. \ref{drbubblefig} is consistent with the GWI.
Consequently, the kernel of the linear response calculated with the
dressed polarization bubble is gauge invariant.

\end{document}